# Accurate Tree Roots Positioning and Sizing over Undulated Ground Surfaces by Common Offset GPR Measurements

Wenhao Luo, *Student Member, IEEE*, Yee Hui Lee, *Senior Member*, *IEEE*, Lai Fern Ow, Mohamed Lokman Mohd Yusof, and Abdulkadir C. Yucel, *Senior Member, IEEE*

*Abstract*— Tree roots detection is a popular application of the Ground-penetrating radar (GPR). Normally, the ground surface above the tree roots is assumed to be flat, and standard processing methods based on hyperbolic fitting are applied to the hyperbolae reflection patterns of tree roots for detection purposes. When the surface of the land is undulating (not flat), these typical hyperbolic fitting methods becomes inaccurate. This is because, the reflection patterns change with the uneven ground surfaces. When the soil surface is not flat, it is inaccurate to use the peak point of an asymmetric reflection pattern to identify the depth and horizontal position of the underground target. The reflection patterns of the complex shapes due to extreme surface variations results in analysis difficulties. Furthermore, when multiple objects are buried under an undulating ground, it is hard to judge their relative positions based on a B-scan that assumes a flat ground. In this paper, a roots fitting method based on electromagnetic waves (EM) travel time analysis is proposed to take into consideration the realistic undulating ground surface. A wheel-based (WB) GPR and an antenna-height-fixed (AHF) GPR System are presented, and their corresponding fitting models are proposed. The effectiveness of the proposed method is demonstrated and validated through numerical examples and field experiments.

## I. INTRODUCTION

The ground-penetrating radar (GPR) is widely applied to the underground detection of tree root systems because of its advantages such as non-destructive, portability, and ease of deployment [1],[2]. Under normal circumstances, a common-offset configured wheel-based (WB) GPR system, also known as ground-coupled GPR, is moved straight over an assumed flat and homogeneous ground, and the radargrams showing hyperbolic features are observed [3-5]. Methods focused on dealing with hyperbolae obtained with these assumed ideal experimental conditions are adopted. One of such methods proposed by W. Li and C. G. Windsor uses the Randomized Hough Transform to determine the hyperbola parameters to recognize hyperbolas in GPR radargrams [6], [7]. A. Abderrahmane applied matched filter technique to the hyperbolic received signal, and then maximizes the signal-to-noise ratio (SNR) at the point corresponding to the target in order to get the location of the target [8]. Jian Chen adopted hyperbolic matching technology to estimate the geological wave velocity and realize depth conversion [9]. All these methods ignore the realistic geographic features of the undulating ground surface. Besides, a large number of approaches, which includes modified gradient method [10], standard or modified Born iterative methods [11], first or higher order Born approximations [12], distorted Born iterative method [13], linear sampling method [14], synthetic aperture technique [15], and genetic or memetic algorithms [16], have been introduced during the last two decades. Most of these approaches consider layered media which contain planar interfaces only.

The surface of the land are naturally not flat and has a soil surface roughness (SSR) [17] due to natural or anthropogenic phenomena, including plant growth, erosion, raindrop impact, and physical crusting, etc. [18], [19]. In the GPR application of tree roots detection, the target in general has a radius of several centimeters. The variation of the air-ground interface will affect the tree root's reflection pattern, turning it into an irregular curve rather than a hyperbola. When multiple roots are buried, it is difficult to determine their relative position because of the different ground height (due to the undulating ground). Therefore, the traditional hyperbola fitting methods cannot be adopted.

In the literature, there are some studies that deal with the detection of objects buried under rough surfaces. In [20], a method based on angular correlation function (ACF) processing, is applied for the detection of a metallic object buried under rough soil surface. In [21], a method based on the correlation of scattered fields from two transmitters is proposed. The methods proposed in [20] and [21] are able to detect buried objects but lack in their ability to accurately determine the shape

Manuscript received December 25, 2021, revised April 19, 2022, accepted May 26, 2022. This work was supported by the Ministry of National Development Research Fund, National Parks Board, Singapore. (*Corresponding authors: Yee Hui Lee; Abdulkadir C. Yucel*).

Wenhao Luo, Yee Hui Lee, and Abdulkadir C. Yucel are with School of Electrical and Electronic Engineering, Nanyang Technological University, 50 Nanyang Ave, 639798, Singapore (e-mail: wenhao.luo@ntu.edu.sg; eyhlee@ntu.edu.sg; acyucel@ntu.edu.sg).

Lai Fern Ow and Mohamed Lokman Mohd Yusof are with National Parks Board, 1 Cluny Rd, 259569, Singapore (e-mail: genevieve_ow@nparks.gov.sg; mohamed_lokman_mohd_yusof@nparks.gov.sg).

and size of the buried objects. In [22], a level-set formulation (LSF) is used to obtain the shape of the buried object provided that the permittivity is known. The surface profile of the soil was taken into account by using the correlation of the scattered field. In [23], a reciprocity gap linear sampling method (RG-LSM) coupled with an analytic continuation method was proposed to localize and determine the shape of the object. However, this method requires the knowledge of the field distributions at the interface. Meanwhile, both these methods require knowledge of the soil surface information and soil characteristics. Both [22] and [23], are unable to determine the shape and size of the buried object accurately even under slightly undulating ground surfaces. The distorted born iterative method (DBIM) was used to obtain both the shape and the properties of the buried object based on updating the Green's function of the layered media with rough interface at each iteration step in [24]. In [25], the buried object approach (BOA) for the determination of the Green's function of a layered media with locally rough interfaces based on assuming the perturbations of the rough surface are buried objects in a layered media with planar interface is proposed. Both these methods rely on the correct selection of many parameters to determine the quality of the reconstruction, and this results in high computational costs.

In the study of tree root detection over undulating ground, we considered two typical kinds of GPR systems. The first is the WB GPR system, a widely used GPR that is efficient and easy to move in the test field [3], [4]. However, when the test terrain is too complex, the antenna-height-fixed (AHF) GPR, an air-coupled GPR, is used to prevent damage to the measurement instruments including the antennas [26], [27]. To take into account the ground undulations so as to determine the accurate position of the underground roots, the GPR system needs a positioning module that initializes a reference height zero ($H_0$). Based on $H_0$ and the test range, a 2-D coordinate system is established. The ground surface is mapped in the coordinates system with coordinate points, and the approximate ground surface below the GPR scanning trace can be obtained through linear interpolation. When the WB-GPR system moves along the terrain, it records the distance it has traveled at each A-scan testing point. Through an arc length approach [28], [29], the distance can be transformed into antenna's coordinates in the established 2-D coordinates system. For the AHF-GPR system, the antennas' height is constantly set at $H_0$, and the antenna horizontal coordinates are the distance from the starting point of the measurement. Once the coordinate of the antenna at each A-scan point is determined, the travel time model of the EM waves corresponding to the two GPR systems are parameterized.

After raw data collection, basic pre-processing methods, such as the time-zero correction method [30] and band-pass filtering technique [31] are applied. Subsequently, singular value decomposition (SVD) is adopted to remove the ringing noise [32], [33]. In order to determine the region of interest (ROI) corresponding to the target accurately and extract the useful time and distance information, a column-connection clustering (C3) algorithm [5], [34] is employed to extract the reflection pattern of each target. Finally, a global optimizer called particle-swarm optimization (PSO) [35], [36] is used to minimize the difference between the modeled travel time and the actual travel time obtained from the extracted radargram. Through this process, the center coordinates and the radii of the roots are determined.

In this paper, a method is proposed for the accurate positioning of the roots and the size of the roots under any realistic terrain. First, we apply a pattern-fitting method to the WB-GPR system. In this case, the positions and radii of the tree roots under the complex-shaped ground can be fitted. Then, the fitting process has been modified accordingly and used in the AHF-GPR system. The estimated results of the tree roots' positions and radii are validated using both synthetic and experimental data.

The paper is organized as follows. Section II discusses the establishment of the testing coordinate system and the determination of the antenna coordinates for every B-scans in both the WB and AHF GPR systems. The EM waves travel time models for the two GPR systems are also introduced. Section III details the proposed method for the root recovery process under complex ground surface terrains from B-scan data. The demonstration of the performance of the two GPR systems and their corresponding fitting model for tree roots recovery are presented in Section IV with simulation data and Section V with measurements data. Section VI concludes the paper.

## II. Basic Concepts and Theories

For GPR systems, the ability to accurately establish the reference position of the GPR will determine the accuracy of the positions of underground objects. Therefore, before fitting the roots based on the reflected patterns, the location of the GPR system in the test scene has to be determined accurately. In the undulated ground scenario, the shape information of the ground surface needs to be taken into consideration. In this section, the antenna coordinate of each scanning point is determined and the ground surface shapes is illustrated. EM waves travel-time model for the two GPR systems with the antenna positions are introduced.

*A. Establishing the Coordinate System and Determining the Antenna Coordinate*

Common offset WB-GPR [3], [4] and AHF-GPR system [26] have been demonstrated to be feasible for survey over flat ground. In that circumstance, a B-scan is an image made up of several A-scans along a straight path over the ground. The B-scan gives an indication of the underground scenario over the testing path. For the flat ground surface, it is treated as the time zero reference point [30], and the position of the target is determined by the travel time of the target's return signal and the EM wave velocity in the ground. Before determining the position of the underground object, the coordinates of the transceivers at all A-scan points are recorded along a straight path.

Over the undulated ground surface, the AHF-GPR system is set and moves at a constant reference height of $H_0$, as shown in the Fig.1. The y coordinate of the antenna for each A-scan is along the straight path, $H_0$. The ground surface (illustrated as the blue curve in Fig.1) coordinates are recorded by the distance measured vertically from $H_0$ at a step of 1 cm along the straight path. Furthermore, interpolating the ground surface coordinates to ensure the abscissa difference between two adjacent points on the surface is 0.1 cm. Through the above process, a 2-D coordinate system is established as in Fig.1. The starting point of the AHF-GPR system is set to be the origin point (0,0) of the 2-D coordinate system.

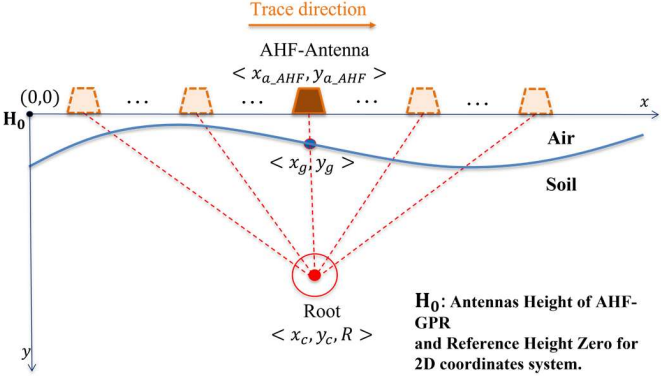

Fig. 1. Established 2D coordinates system and AHF-GPR data collection in the curved ground scenario.

For the WB-GPR, the GPR is moved along the undulating ground surface rather than a fixed height above ground in Fig.1. Inspired by the work of Giannakis and Tosti used for tree trunk health monitoring [37], [38], arc-length parameterization [28], [29] is used here to convert the system's travelling distance to 2-D coordinates. A WB GPR system with traveling distance measurement is very common in commercial usage [37], [38]. Based on the $H_0$ and the ground surface coordinates we recorded in the process of 2D coordinates establishment in the last paragraph, we can get the shape of the ground surface, which is made up with coordinates sets $\{x, y \in \mathbb{R} \mid x, y > 0\}$ of $n$ points along the test route, where $x$ is the horizontal position and $y$ is the vertical distance from the set height $H_0$. For parameterizing the complex-shape ground surface by a single variable, we map the coordinates to the variable $\{u \in \mathbb{R}^n \mid u \in [0,1]\}$ by spline interpolation. After that, the surface shape can be represented as $\mathbf{S}(u) = (x(u), y(u))$. GPR Route along the soil surface, which can be assumed as the distance the WB-GPR travels, is denoted by $s$.

$$s(\mu) = \int_0^\mu \left\| \frac{d\mathbf{S}}{du} \right\| du = \int_0^\mu \sqrt{\left(\frac{dx(u)}{du}\right)^2 + \left(\frac{dy(u)}{du}\right)^2} \, du \quad (1)$$

The integral in the equation can be approximated numerically by the summation of relative distance for each step. Then, a spline interpolation is used to map the $\mu$ with respect to $s$.

Upon above, once the distance $s$ traveled by the WB-GPR system from the starting point to a certain A-scan test position is recorded, we can get the corresponding coordinates of its location, that is, the antenna coordinates.

Through the above process, a 2-D coordinate system is established, where the starting point of the AHF-GPR is used as the origin point (0,0). The antenna coordinates and the ground shape are recorded.

*B. EM Waves Travel-Time Model of the Two GPR Systems*

To accurately obtain the position and size of the cylindrical object through B-scan, we need to take the geographic feature of the non-planar ground into consideration.

In [39] and [40], when the host medium is a cylinder, the scheme that describes the reflection pattern of the internal objects is proposed. In our scenario, it can be described by the following equations:

$$t_{WB} = \left( \sqrt{(x_{a\_WB} - x_c)^2 + (y_{a\_WB} - y_c)^2} - R \right) \frac{2\sqrt{\varepsilon}}{c_0} \quad (2)$$

$$t_{AHF} = t_{air\_AHF} + t_{soil\_AHF} \quad (3)$$

where

$$t_{air\_AHF} = \left( \sqrt{(x_{a\_AHF} - x_g)^2 + (y_{a\_AHF} - y_g)^2} \right) \frac{2}{c_0} \quad (4)$$

$$t_{soil\_AHF} = \left( \sqrt{(x_g - x_c)^2 + (y_g - y_c)^2} - R \right) \frac{2\sqrt{\varepsilon}}{c_0} \quad (5)$$

$t_{WB}$ and $t_{AHF}$ are the modelled travel time of EM wave in the circumstances of WB and AHF GPR respectively. $t_{air\_AHF}$ is the time that EM wave transmitted in the air and $t_{soil\_AHF}$ is the time that EM wave transmitted in the soil when the AHF-GPR is applied. $\varepsilon$ is the relative permittivity of the soil, $c_0$ is the speed of light in air ($c_0 \approx 3 \times 10^8 \, m/s$). $<x_{a\_WB}, y_{a\_WB}>$ are the coordinates of the antennas along the surface of the ground for the WB-GPR in equation (2); and $<x_{a\_AHF}, y_{a\_AHF}>$ are the coordinates of the antennas along the scanning trace of the AHF-GPR in equation (4). $y_{a\_WB}$ is the relative height with respect to $H_0$ mentioned in Section II-A and $y_{a\_AHF}$ is at the same height of $H_0$, shown in Fig. 1. $<x_c, y_c>$ are the coordinates of the roots center while R is the radius of the root. $<x_g, y_g>$ are the coordinates where the EM wave hits the surface when the antennas are lifted above the surface in the AHF-GPR system. The travel-time model for AHF-GPR uses the rectilinear transmission assumption of the electromagnetic wave [22], [24], as illustrated in Fig. 1. The distance between the antennas and the soil surface is always in the near-field range. This helps to ensure that the working status of the antennas is not disturbed, energy is not wasted in the air, the responses of small targets are substantially kept, and other effects due to ground surface obstructions that are not directly below the system are not introduced [41].

## III. DATA PROCESSING

The data obtained from numerical and field measurements need to be processed to get the target's position and size. First, the data undergoes preprocessing to reduce the ringing noise resulting from the nature of the complex heterogeneous soil. Then the regions of interest (ROI) in the B-scan after preprocessing are extracted to get the reflection-arrival travel time corresponding to each target. The difference between the actual travel time (from radargrams) and the theoretical travel time (given in equation (2) and (3)) is minimized, in order to approximate the center position and radius of the tree root. The data processing is implemented in MATLAB R2020b. The workstation used for data processing has Intel(R) Xeon(R) W-2255 CPU @ 3.7 GHz processor, 64 GB RAM, and NVIDIA Quadro P400 GPU.

*A. Pre-processing methods and ROI extraction*

The raw data obtained by GPR need to be processed and interpreted carefully. The starting reference time of the GPR data is fixed by a time-zero correction method [30]. The signal-to-noise ratio (SNR) is improved by filtering out the signal components with frequencies outside the operating bandwidth through the band-pass filtering technique [31]. Horizontal clutter due to the direct coupling between antennas and reflection from the ground surface is suppressed via a background removal method: singular value decomposition (SVD) [32], [33] which keeps the intermediate eigenvalues while setting others to zero.

The preprocessing above helps to improve the SNR in most testing scenarios. To achieve the goal of recovering the position and the size of the roots, we need to focus on the ROI corresponding to each root in the B-scan. For that purpose, we adopt the C3 algorithm [5] [34]. In the C3 algorithm, we first treat a set of adjacent pixels in each column of the B-scan as a segment. Then from the first to the last column of the B-scan, adjacent segments in the connected columns are clustered together as a 'region' to form the ROI. This helps to differentiate the ROI and the noise. Several thresholds, such as the number of pixels of a segment, shared pixels of two connecting segments in two columns, and the total number of pixels of a 'region' are set in order to reduce the noise. Finally, by extracting the pixel coordinates and corresponding values of ROIs in the original B-scan, each ROI is extracted. The details of how the C3 algorithm works are illustrated in [34].

*B. Proposed Technique for Targets Position and Size Retrieval*

In this section, the roots' radii and center positions, illustrated in the travel time equations (2) to (5), are determined. The reflection patterns of roots change with the undulating ground surfaces. Due to the undulating ground surface, the reflection patterns from the tree roots are not a hyperbola. Therefore, in our proposed technique the method used here takes that into consideration and is more practical.

Through the C3 technology, the ROI can be extracted, and the travel time $t_a$ corresponding to each antenna position $<x_a, y_a>$ can be obtained by the methods in Section II.A. $P = \{(x_a, y_a, t_a) \in \mathbb{R}, |a = 1, 2 \cdots z\}$ is known and defined to represent ROI, where $z$ is the number of observations. Then, we defined two cost functions

$$\sum_{a=1}^{z}\left(t_a - \left(\left(\sqrt{(x_{a\_WB} - x_c)^2 + (y_{a\_WB} - y_c)^2} - R\right)\frac{2\sqrt{\varepsilon}}{c_0}\right)\right)^2 \quad (6)$$

and

$$\sum_{a=1}^{z}\left(t_a - (t_{air\_AHF} + t_{soil\_AHF})\right)^2 \quad (7)$$

corresponding to (2) and (3), respectively. $t_{air\_AHF}$ and $t_{soil\_AHF}$ are the same as that in (4) and (5). The two cost functions are subjected to the following parameters: center coordinates of target $<x_c, y_c>$, the radius of the target $R$ and relative permittivity of the soil $\varepsilon$. Once the permittivity of the soil is known, by substituting $P$ into (6) and (7), and minimizing the cost functions, the center position of the root $<x_c, y_c>$, and the radius $R$ of the root can be approximated. This can be done for both the WB-GPR system and AHF-GPR system.

We need to emphasize that, in order to optimize equation (7), the surface shape of the soil should be known. The coordinates information of the soil surface is collected by the position module of the GPR system, The specific implementation process is described in detail in Section II A.

In this paper, PSO, a popularly used global optimizer [35], [36], is chosen to minimize the two cost functions. PSO is an optimizer that is modeled after the social behavior of swarms of insects. Initially, a number of candidate solutions (or particles) $N$ are randomly group together in the search space. In this case, if the permittivity is known, each candidate solution is represented as $q_k = \langle x_k, y_k, R_k \rangle \{q_k \in \mathbb{R}^3, k = 1,2,...,N\}$. The elements of the solutions are used to evaluate the cost functions and are adjusted iteratively in the search space according to the rule in (8).

$$v_k^\tau = \varphi_0 v_k^{\tau-1} + \varphi_1 V_1(q_k^{\tau-1} - q_{k,b}) + \varphi_2 V_2(q_k^{\tau-1} - q_g) \quad (8)$$

$$q_k^\tau = q_k^{\tau-1} + v_k^\tau \quad (9)$$

where $\varphi_0$, $\varphi_1$, and $\varphi_2$ are learning rates governing the procedure of finding the minimal solutions. The parameters $V_1$, $V_2 \in [0, 1]$ are random numbers with a uniform distribution. The vector $v_k \in \mathbb{R}^3$ contains the gradient (direction and speed) that $q_k$ would travel. Corresponding to all candidates, the cost function has values $C_k\{k = 1,2,...,N\}$ at the beginning of the process. In each iteration $\tau$ ($\tau \in Z$), the k[th] candidate adjusts itself to $q_{k,b}$, a new root center position and size, and the cost function values corresponding to every particle get their minimum values $C_{k\_min}\{k = 1,2,\cdots,N\}$. In this iteration, the cost function achieves its minimal value ($min\left(C_{k\_min}\{k = 1,2,...,N\}\right)$), and the related $<x, y, R>$ is the vector $q_g$. In our case, the parameters: $N = 100$ and $\varphi_0 = 0.5$, $\varphi_1 = -1.5$, $\varphi_2 = -1$ are derived with both efficiency and accuracy.

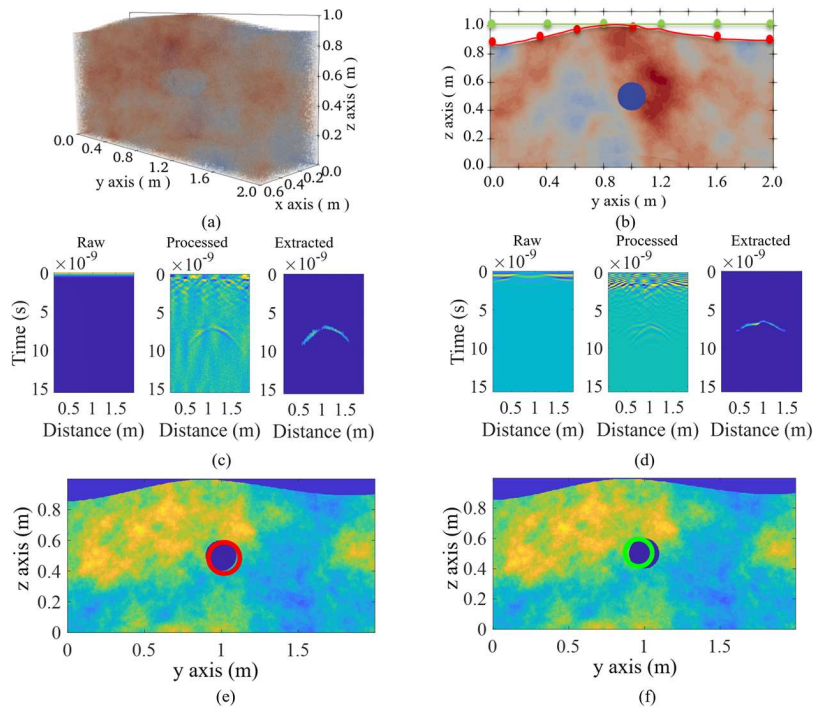

Fig. 2. (a) Three-dimensional scenario model of one root buried under non-planar ground. (b) Two-dimensional slice of the scenario. Red curve: scanning trace of the WB-GPR. Green line: scanning trace of the AHF-GPR. (c) Process of the B-scan of the WB-GPR. (d) Process of the B-scan of the AHF-GPR. (e) Red circle: recovered position and radius of the root by WB-GPR. (f) Green circle: recovered position and radius of the root by AHF-GPR.

## IV. Numerical Experiment

In the numerical experiments, three scenarios where tree roots are buried in three different undulating soil surfaces are simulated. The ability of the two types of GPR systems and their corresponding fitting models on detecting tree roots is investigated.

The numerical simulation of tree roots in a heterogeneous soil environment is carried out by an open-source gprMax software [42]. In gprMax, the finite-difference time-domain technique is used to characterize the transient EM phenomena. In an effort to accelerate the computation, an open-source compute unified device architecture (CUDA)-based GPU engine [43] is used. The roots have a volumetric water content of 50% with a relative permittivity of $\varepsilon_{root} = 24$ and electrical conductivity of $\sigma_{root} = 0.63$ mS/m for numerical simulation purposes [44], [45]. The magnetic permeability and magnetic conductivity are set to 1 and 0, respectively.

Parameters of heterogeneous soil are set by using the existing Peplinski model which leverages on the semi-empirical model [46], [47] and the fractal model [48] in gprMax. In the simulation, the soil has a sand fraction ($S$) of 0.3, a clay fraction ($C$) of 0.7, a sand particles density ($\rho_s$) of 2.66 gr/cm$^3$, a bulk density ($\rho_b$) of 2 gr/cm$^3$, and a water volumetric fraction ($f_u$) varying from 0.01 to 0.15. The number of soil elements and the fractal dimension ($\beta$) are set to 20 and 1.5, respectively. The average relative permittivity of the soil is equal to 6.02, which is used as a known constant $\varepsilon$ in the equation (2) and (5) for simulation cases.

The first two case studies are shown in Fig. 2 and Fig. 3. One cylindrical root with a radius and length of 0.1 m and 0.3 m, respectively is buried under two different undulated ground surfaces. In the third case study, three roots with different radius (0.15 m, 0.10 m, 0.08 m) are buried under different depths of another undulated ground surface as shown in Fig. 4. The size of the soil box is 0.6 m by 2 m, with a depth of 1 m. The spatial spacing of the different grids in the numerical calculation is set to 0.002 m.

For the excitation of the FDTD, a numerical equivalent of the commercial antenna GSSI 1.5 GHz is used in the AHF GPR system [49], [50]. The excitation of the antenna is parallel to the orientation of the root (x-axis) in order to maximize the reflected signal of an elongated object. The excitation is placed at a constant height of 0.01 m above the highest point of the undulated ground surface along the scanning trace. 85 A-scans are recorded for each case from 0.12 m to 1.80 m along the y-axis with a step of 0.02 m. The scanning traces of the AHF-GPR system are represented by the green lines in Fig. 2(b), Fig. 3(b), and Fig. 4(b).

The excitation source chosen for the WB-GPR study is an ideal Hertzian dipole using a modulated Gaussian pulse with a 1.0 GHz central frequency. Note that the antenna GSSI model in gprMax can only be parallel to one of the Cartesian axes, it cannot be placed at other angles. Due to the size of the GSSI model, there will be collisions between the antenna model and the ground. The ground surface will limit the placement of the antenna GSSI on the ground. The scanning trace follows the surface of the undulated ground surface along the y-axis and is represented by the red curve line in Fig. 2(b), Fig. 3(b), and Fig. 4(b). The A-scan scanning steps for both cases is 0.02 m. Since the curve lengths of the three ground surface cases are different, a total of 86, 101, and 90 A-scans are recorded for the three numerical cases, along their ground surface.

In our simulation, the highest operating frequency is 3.1 GHz in gprMax when the center frequency is set as above, the calculated wavelength is approximately 0.03 m for soil permittivity of around 6 to 8. The roots' diameters we simulated are bigger than the wavelength, and the system settings are theoretically feasible.

Fig. 2, 3, 4 (c) and (d), shows the resulting raw B-scans obtained from WB and AHF GPR systems in the three scenarios respectively. All the B-scans are processed using the SVD filter to remove the three and six dominant eigenvalues for WB and AHF GPR systems, respectively. The features that indicate the roots are then extracted by using the C3 technique. The raw data, preprocessed data, and extracted parabola are shown in Fig.2(c)(d). It is worth noting that the reflection patterns extracted in from an undulating ground, is more complex and far more distorted as compared to that from a typical GPR over a flat ground surface where the reflected pattern would be a clear parabola. This makes the patterns from an undulated ground surface difficult to interpret. If the specific position of the tree root is determined according to the highest point or inflection point of the patterns as is typically done, there will be significant error. The data processing is omitted in Fig. 3(c)(d) and Fig. 4(c)(d), and the extracted patterns are shown.

From Fig. 2(e) (f), Fig. 3(e) (f), and Fig. 4(e) (f), it is evident that the proposed method can recover both the position and radius of the roots accurately after considering the geographic feature of the undulating ground surface of the test sites.

The estimated (*est*) and real (*real*) results of the coordinate positions and radius of the roots in the three scenarios are summarized in Table I, II. From the tables, we can see that both AHF and WB GPR systems can predict the position and size of the roots efficiently using our proposed technique. The accuracy is both high and slightly higher for the WB as compared to the AHF.

The computational time taken by the proposed method to process the three numerical experiences are displayed in Table III as $t$. From Table III, we can see that the computational time for AHF-GPR is more than WB-GPR. This is because, equation (3) and (7) has more complexity than equation (2) and (6). Furthermore, the coordinates $<x_g, y_g>$ where the EM wave hits the surface are determined automatically in the process by finding the intersection of EM transmission path and soil surface curve, which increases the computational time of the AHF-GPR system.

The LSF method in [22] is compared with our methods. Both the LSF and our proposed methods uses bistatic measurement. In [22], a distance function $L_1$ is used to determine the accuracy of the reconstructed shape and the actual shape. The simulated results are compared and their $L_1$ values and average cost time per radargram (one ROI) are listed in the TABLE III.

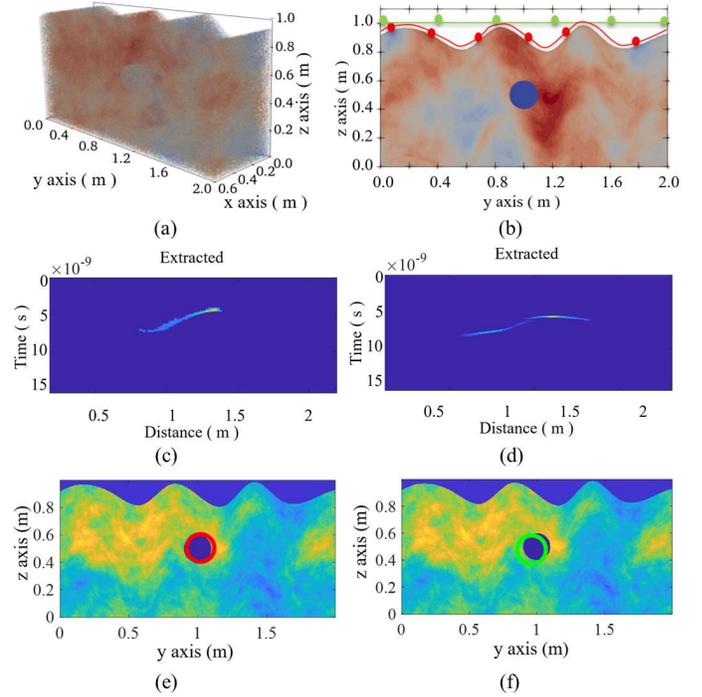

Fig. 3. (a) Three-dimensional scenario model of one root buried under extreme non-planar ground. (b) Two-dimensional slice of the scenario. Red curve: scanning trace of the WB-GPR. Green line: scanning trace of the AHF-GPR. (c) The extracted pattern of the WB-GPR B-scan. (d) The extracted pattern of the AHF-GPR B-scan. (e) Red circle: recovered position and radius of the root by WB-GPR. (f) Green circle: recovered position and radius of the root by AHF-GPR.

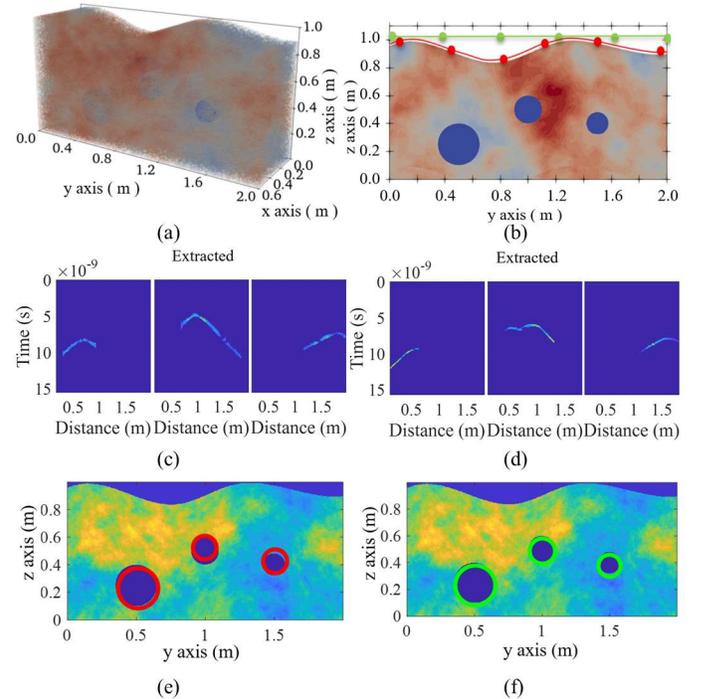

Fig. 4. (a) Three-dimensional scenario model of three roots buried under non-planar ground. (b) Two-dimensional slice of the scenario. Red curve: scanning trace of the WB-GPR. Green line: scanning trace of the AHF-GPR. (c) The extracted patterns of the WB-GPR B-scan. (d) The extracted patterns of the AHF-GPR B-scan. (e) Red circle: recovered position and radius of the root by WB-GPR. (f) Green circle: recovered position and radius of the root by AHF-GPR.

Comparing the shape reconstruction results of LSF with the results obtained by our proposed methods, it is evident that:

1) The roots reconstruction results of the LSF method are worse than the proposed method. This is due to the existence of environmental noise and clutter in the heterogeneous soil simulation. As stated in [22], the $L_1$ results of LSF method is vulnerable to noise and clutter.
2) The average time of the LSF method is about 10 mins, which is about 20 times more than the WB-GPR fitting method and 6 times more than the AHF-GPR fitting method.

TABLE I
RADIUS ($r$ (m)) ESTIMATION USING THE AHF AND WB GPR SYSTEM WITH THEIR CORRESPONDING FITTING MODELS

| Scenarios (S) | S1 | S2 | S3 | | |
|---|---|---|---|---|---|
| | | | Root1 | Root2 | Root3 |
| WB GPR system | | | | | |
| $r_{real}$ | 0.100 | 0.100 | 0.150 | 0.100 | 0.080 |
| $r_{est}$ | 0.102 | 0.101 | 0.145 | 0.085 | 0.086 |
| AHF GPR system | | | | | |
| $r_{real}$ | 0.100 | 0.100 | 0.150 | 0.100 | 0.080 |
| $r_{est}$ | 0.090 | 0.106 | 0.148 | 0.094 | 0.082 |

TABLE II
CENTER POSITION ($C$ (m, m)) ESTIMATION USING THE AHF AND WB GPR SYSTEM WITH THEIR CORRESPONDING FITTING MODELS

| Scenarios (S) | S1 | S2 | S3 | | |
|---|---|---|---|---|---|
| | | | Root1 | Root2 | Root3 |
| WB GPR system | | | | | |
| $C_{real}$ | (1.00, 0.50) | (1.00, 0.50) | (0.50, 0.25) | (1.00, 0.50) | (1.50, 0.40) |
| $C_{est}$ | (1.02, 0.49) | (1.02, 0.51) | (0.51, 0.23) | (0.99, 0.52) | (1.51, 0.42) |
| AHF GPR system | | | | | |
| $C_{real}$ | (1.00, 0.50) | (1.00, 0.50) | (0.50, 0.25) | (1.00, 0.50) | (1.50, 0.40) |
| $C_{est}$ | (0.96, 0.51) | (0.96, 0.48) | (0.50, 0.23) | (1.00, 0.49) | (1.49, 0.38) |

TABLE III
$L_1$ DISTANCE BETWEEN THE RECONSTRUCTED ROOTS AND THE REAL ROOTS BY DIFFERENT METHODS, AND THEIR AVERAGE COST TIME FOR SIMULATION.

| Method \ Scenarios (S) | S1 | S2 | S3 | | | $t$ (s) |
|---|---|---|---|---|---|---|
| | | | Root1 | Root2 | Root 3 | |
| WB | 0.28 | 0.30 | 0.62 | 0.12 | 0.46 | 37 |
| AHF | 0.35 | 0.37 | 0.63 | 0.33 | 0.31 | 155 |
| LSF | 0.67 | 0.55 | 1.35 | 0.40 | 0.73 | 660 |

## V. FIELD EXPERIMENT

Field experiments are carried out to validate the effectiveness of the WB and AHF GPR systems and their corresponding proposed fitting methods. The testing scenarios are shown in Fig. 5. To clearly show the effectiveness of the method, the cylindrical subsurface targets in the experiment are three freshly collected tree roots with high water content and therefore good electromagnetic reflectivity. In Fig. 5(a), the tree roots have radii around 3.10 cm, 2.50 cm, and 1.75 cm.

A Keysight P5008A Vector Network Analyzer (VNA) is used as a transceiver. It sweeps 1001 frequency points within the frequency band of 0.4 GHz to 3.4 GHz. The intermediate frequency bandwidth (IFBW) is set as 500 Hz [51], and the power is set as -10 dBm. The setup of the WB-GPR and AHF-GPR systems are shown in Fig. 5(b) and (c), respectively. Vivaldi antennas described in [52] are arranged in a quasi-monostatic configuration placed 5 cm apart from each other, are used as transmitter and receiver antennas. The antennas are oriented parallel to the direction of the roots. In the WB-GPR system, the antennas are placed in a foam box with wheels. A distance recording module is connected to the wheel similar to those used in a commercial GPR system [37], [38]. The system is then moved along the undulated ground surface. The size and the configuration of the system are appliable for measurements over curved ground surfaces. In the AHF-GPR system, the foam box is raised to a fixed height of $H_0$ by a wooden frame and moved at a constant height over the undulated surface.

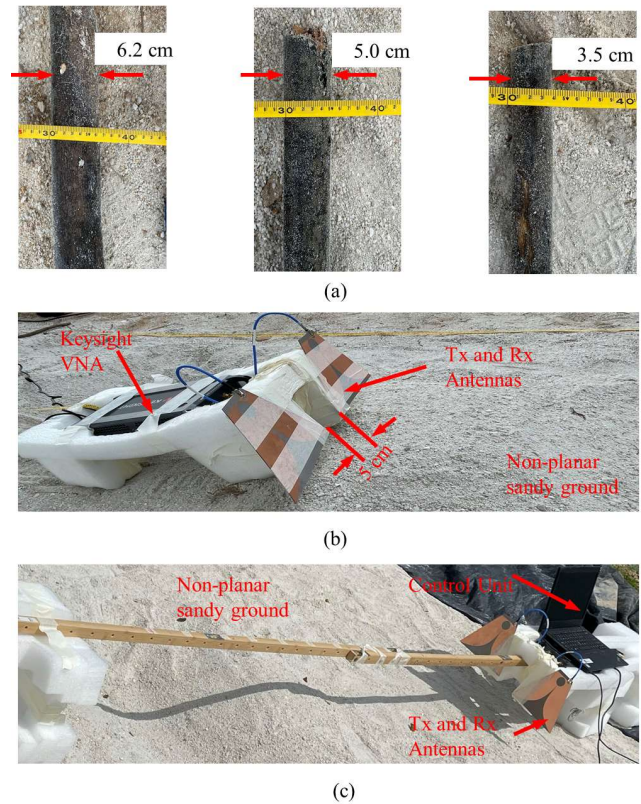

Fig. 5. The testing scenes. (a) Three tree roots used as targets in the experiment. (b) WB-GPR system set up. (f) AHF-GPR system set up.

The measurement scenario is illustrated in Fig. 6 (a), (b). An undulated ground surface is built where the biggest height difference of the undulated ground surface is 0.21 m in height. The horizontal distance between the highest point and the lowest point is 0.56 m apart. The fixed height of the AHF-GPR system, $H_0$ is set as 0.16 m above the highest point of the undulated ground surface. The three tree roots are buried under the undulated ground surface. The center positions of the tree roots in the height and distance apart in the coordinate system of Fig. 6 (b) are 0.55 m along the measured path and 0.50 m below the reference height $H_0$, 1.05 m along the measured path and 0.25 m below the reference height $H_0$, and 1.55 m along the measured path and 0.40 m below the reference height $H_0$, respectively.

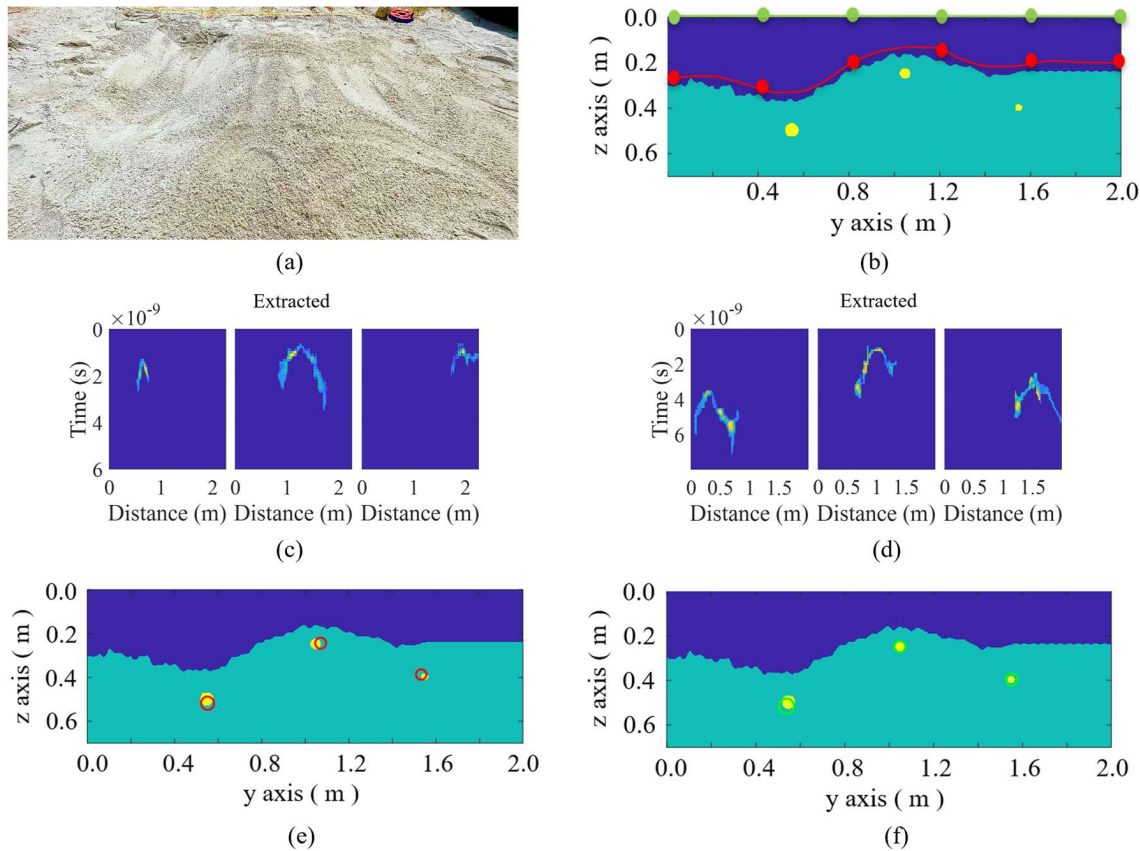

Fig. 6. (a) Sandy test site with non-planar surface. (b) Two-dimensional slice of the scenario. Red curve: scanning trace of the WB-GPR. Green line: scanning trace of the AHF-GPR. (c) The extracted patterns of the WB-GPR B-scan. (d) The extracted patterns of the AHF-GPR B-scan. (e) Red circle: recovered position and radius of the root by WB-GPR. (f) Green circle: recovered position and radius of the root by AHF-GPR.

First, we examine the capability of the WB-GPR system and its corresponding method in fitting the tree roots' positions and sizes. The red curve in Fig. 6(b) shows the trace path that the GPR system follows. A total of 100 A-scans are recorded every 2 cm interval along the path. Using the approach in Section II, the travel distance is transformed to coordinates. The raw data is pre-processed through zero-time removal, time-varying gain, and direct current (DC) removal. Subsequently, an SVD and C3 technique are applied to reduce the ringing noise and perform pattern extraction over the region of interest. For this experiment, three dominant eigenvalues are filtered out in SVD, and three patterns corresponding to the three roots are extracted.

Fig. 6(c) shows the extracted reflection patterns of the roots. The label for abscissa is the distance that the system travels. Given the position of the three radargrams in the B-scan, we cannot tell the relative position of the three roots underground directly by the B-scan. Specifically: the three radargrams are at the same time section in the B-scan, but they are actually with different relative position in the coordinate system. In order to estimate the position/size of the roots that fit these patterns, the relative soil permittivity over the targets is estimated at different depth intervals and horizontal positions with a step of 0.25 m in the range of test field. The relative soil permittivity data is collected using the 85070E dielectric probe. The averaged value of the relative soil permittivity ($\varepsilon$) is estimated to be 5.2. The corresponding depth resolution of the set frequency band is:

$$\Delta\delta = \frac{1}{2 \times BW} \cdot \frac{c_0}{\sqrt{\varepsilon}} \qquad (10)$$

$c_0$ is the speed of EM wave in air. $BW = 3$ GHz is the bandwidth of the system, the calculated depth resolution is 0.02 m. The depth resolution is small relative to the diameters of the roots we buried, which means the settings of the system are appropriate. The GPR antenna parameters (centre frequency, bandwidth, etc.) and root parameters (diameter, depth, etc.) affect the measurement results, the antenna parameters we selected in this work are based on the medium characteristics, roots size, and target depth. The system setting works well in our scenario. To detect thinner roots, researchers are recommended to increase the bandwidth of the system within antennas' working band limitation. To detect roots in deeper soil, researchers should make sure the higher frequencies can reach that depth and return effective signals. Centre frequency should be adjusted according to the target depth and the medium characteristics.

The recovered roots using the WB GPR system are shown in Fig. 6(e). Based on the shape of the ground surface and its mean relative permittivity, the minimization in function (6) converged to three targets that best fit the data according to the travel time of the three extracted patterns' EM reflection. It is evident from Fig. 6(e) that the positions, as well as the radii of the roots, can be successfully recovered. The recovered roots center depth and relative position coordinates are listed in Table IV.

Second, the estimation of three roots' position and radii by the AHF-GPR system is also examined. The detection track of the system is shown by the green line in Fig. 6(b). The B-scan is also made up of a 100 A-scans by moving the system along the holding frame with step size of 2 cm. After a similar processing scheme to the WB-GPR system, three features are clearly extracted in Fig. 6(d), the abscissa here is the distance that the system moves. It is obvious that the radargrams of the corresponding tree roots obtained by the two GPR systems are different.

TABLE IV
RADIUS ($r$ (m)) AND CENTER POSITION ($C$ (m, m)) ESTIMATION USING THE AHF AND WB GPR SYSTEM WITH THEIR CORRESPONDING FITTING MODELS

|  | Root1 | Root2 | Root3 |
| --- | --- | --- | --- |
| WB GPR system | | | |
| $r_{real}$ | 0.031 | 0.025 | 0.017 |
| $r_{est}$ | 0.030 | 0.025 | 0.023 |
| $C_{real}$ | (0.55,0.50) | (1.05,0.25) | (1.55,0.40) |
| $C_{est}$ | (0.55,0.51) | (1.07,0.24) | (1.53,0.38) |
| AHF GPR system | | | |
| $r_{real}$ | 0.031 | 0.025 | 0.017 |
| $r_{est}$ | 0.033 | 0.024 | 0.025 |
| $C_{real}$ | (0.55,0.50) | (1.05,0.25) | (1.55,0.40) |
| $C_{est}$ | (0.54,0.51) | (1.05,0.24) | (1.55,0.39) |

TABLE V
$L_1$ DISTANCE BETWEEN THE RECONSTRUCTED ROOTS AND THE REAL ROOTS BY DIFFERENT METHODS, AND THEIR AVERAGE COST TIME FOR FIELD TEST

| Scenarios (S) Method | $L_1$ distance | | | $t$ (s) |
| --- | --- | --- | --- | --- |
| | Root1 | Root2 | Root3 | |
| WB | 0.61 | 0.37 | 0.35 | 34 |
| AHF | 0.72 | 0.16 | 0.07 | 109 |
| LSF | 1.62 | 0.92 | 0.79 | 600 |

By taking the travel time of each extracted radargram, together with the geographic feature of the ground surface and the system height into the equation (7), we can recover three root centers as the results shown in Table IV. There are inevitable errors during the experiment, such as distance measurement error, the slight offset of the positions of the roots, and the medium's relative permittivity measurement error, etc. Despite the errors, it is satisfactory to obtain such a small difference between the estimated result (*est*) and the ground truth (*real*). This verifies our proposed technique for position and size estimation of underground objects over undulating ground using experimental data.

The computational time taken by the proposed processing framework to process the field measurement is given in Table V as *t*. In the field test, we kept the number of A-scans and the number of samples contained in each A-scan at about the same amount as the simulation data, so the computational time of each method in field test is similar to the simulation result.

The field experiments comparison between the LSF [22] and our proposed method is listed in TABLE V. Again, our proposed methods outperform the LSF method. The real measured data has significant amount of noise and clutter and therefore, the LSF method is found to perform worse compared to the simulated results. The average computation time of the LSF methods is significantly higher than the proposed methods.

VI. CONCLUSION

Wheel-based (WB) GPR and antenna height fixed (AHF) GPR systems and their corresponding fitting models are adopted in this paper to derive tree roots under undulating or even complex-shaped ground surfaces. The processing flow is described in detail. Three numerical cases where roots with different radii buried in distinct scenarios are simulated and dealt with. The fitted and the real values corresponds well in terms of the center coordinates and radii of the roots by both the GPR systems. In addition, an experimental result is presented in which both the WB GPR and AHF GPR model successfully recover three roots with different sizes and buried depths under an uneven ground. In view of the accuracy of the results, the proposed technique allows the two different types of GPR to estimate the size and location of tree roots under the different depths of undulated ground surfaces, without being restricted by topography.

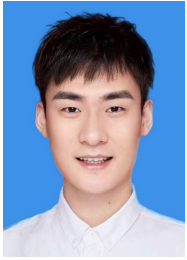

**Wenhao Luo** (Student Member, IEEE) received the B.Eng. degree in automation from Hohai University, Nanjing, China, in 2016, and the M.Sc. degree in electronics engineering in 2019 from the School of Electrical and Electronics Engineering, Nanyang Technological University, Singapore, where he is currently working toward the Ph.D. degree in engineering. His research interests include the ground penetrating radar, remote sensing techniques, radar signal processing and machine learning techniques.

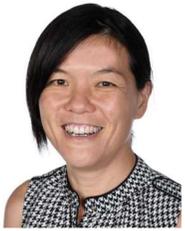

**Yee Hui Lee** (Senior Member, IEEE) received the B.Eng. (Hons.) and M.Eng. degrees from the School of Electrical and Electronics Engineering, Nanyang Technological University, Singapore, in 1996 and 1998, respectively, and the Ph.D. degree from the University of York, York, U.K., in 2002. Since 2002, she has been a Faculty Member with Nanyang Technological University, where she is currently an Associate Professor with the School of Electrical and Electronic Engineering. Her research interests include the channel characterization, rain propagation, antenna design, electromagnetic bandgap structures, and evolutionary techniques.

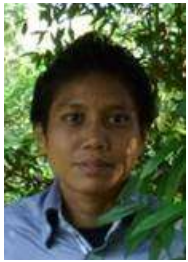

**Lai Fern Ow** received the B.S. degree (Hons.) in plant biology from Massey University, New Zealand, and the Ph.D. degree in plant biology from University of Canterbury, New Zealand, in 2009. She is currently the deputy director with the Centre for Urban Greenery & Ecology, National Parks Board, Singapore. Her research interests include ecophysiology of resource acquisition in urban ecosystems, tree biology/arboriculture, urban sustainability, plant physiology, and responses of plants to extremes of environment.

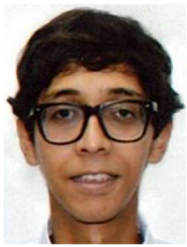

**Mohamed Lokman Mohd Yusof** received the B.S. degree in biomedical science from the University of Western Australia, Crawley, WA, Australia. He is presently an Executive with the Centre for Urban Greenery and Ecology, National Parks Board, Singapore. His research interests include arboriculture, remote sensing, and phytoremediation.

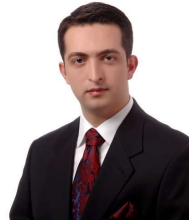

**Abdulkadir C. Yucel** received the B.S. degree in electronics engineering (Summa Cum Laude) from Gebze Institute of Technology, Kocaeli, Turkey, in 2005, and the M.S. and Ph.D. degrees in electrical engineering from the University of Michigan, Ann Arbor, MI, USA, in 2008 and 2013, respectively.

From September 2005 to August 2006, he worked as a Research and Teaching Assistant at Gebze Institute of Technology. From August 2006 to April 2013, he was a Graduate Student Research Assistant at the University of Michigan. Between May 2013 and December 2017, he worked as a Postdoctoral Research Fellow at various institutes, including the Massachusetts Institute of Technology. Since 2018, he has been working as an Assistant Professor at the School of Electrical and Electronic Engineering, Nanyang Technological University, Singapore.

Dr. Yucel received the Fulbright Fellowship in 2006, Electrical Engineering and Computer Science Departmental Fellowship of the University of Michigan in 2007, and Student Paper Competition Honorable Mention Award at IEEE International Symposium on Antennas and Propagation in 2009. He has been serving as an Associate Editor for the International Journal of Numerical Modelling: Electronic Networks, Devices and Fields and as a reviewer for various technical journals.